\documentclass[twocolumn,aps,prl,showpacs,groupedaddress,amsfonts,amssymb,amsmath]{revtex4-1}
\usepackage{graphicx}
\usepackage{xcolor, soul}
\sethlcolor{cyan}
\usepackage{subfigure}
\usepackage{epstopdf}
\usepackage[colorlinks=true, letterpaper=true, pdfstartview=FitV, linkcolor=blue, citecolor=blue, urlcolor=blue]{hyperref}

\begin{document}
\draft

%\title{Co-exitance of quantum anomulas Hall effect and quantum spin Hall effect in the presence of exchange field}

\title{Reentrant quantum spin Hall effect in the presence of an exchange field}

\author{Xiaoyu Cheng,$^{1,4}$ Xiaohong Zheng,$^{2}$ Jun Chen,$^{3,4,*}$ Liantuan Xiao,$^{1,4}$ Suotang Jia,$^{1,4}$ Lei Zhang,$^{1,4,\dagger}$}
\address{$^1$State Key Laboratory of Quantum Optics and Quantum Optics Devices, Institute of Laser Spectroscopy, Shanxi University, Taiyuan 030006, China\\
$^2$College of Information Science and Technology, Nanjing Forestry University, Nanjing 210037, China\\
$^3$State Key Laboratory of Quantum Optics and Quantum Optics Devices, Institute of Theoretical Physics, Shanxi University, Taiyuan 030006, China\\
$^4$Collaborative Innovation Center of Extreme Optics, Shanxi University, Taiyuan 030006, China}

\begin{abstract}
Driven by various physical origins, the interesting reentrant phenomena in quantum Hall effect (QHE), quantum anomalous Hall effect (QAHE) and non-Hermitian systems have been discussed recently. Here, we present that the reentrant phenomena of quantum spin Hall effect (QSHE) can be achieved by introducing a uniform exchange field into a topological insulator with a four-band model Hamiltonian. Furthermore, it is found that both time-reversal (TR) symmetry broken QSHE and spin-polarized QAHE can simultaneously be generated in the system with proper exchange field strength by inspecting the corresponding edge states. Through analyzing the energy dependent spin-Chern number $\nu_s$ and transmission in the presence of random disorders, we verify the topological protection and the robustness of induced QSHE and QAHE. Finally, the theoretical realization of reentrant QSHE in a topolectrical circuit is demonstrated. Our work not only extends the reentrant phenomena into the QSHE but also stimulates interest in their realization in different systems.
\end{abstract}

% PACS number(s):

\maketitle

\emph{Introduction}-As one of the most promising potential low-dissipation candidates in spintronic devices, the quantum spin Hall effect (QSHE) has attracted a lot of research attention on both theoretical and experimental sides\cite{murakami2003dissipationless,kane2005quantum,sinova2004universal,kato2004observation,wunderlich2005experimental,bernevig2006quantum,kane2005z,sheng2005nondissipative,sheng2006quantum,murakami2006quantum,xu2006stability,sheng2005nondissipative,bernevig2006quantum2,qi2006topological,wu2006helical,onoda2007localization,konig2008quantum,ok2019custodial,lee2021gate,konig2007quantum,nowack2013imaging,deacon2017josephson,liu2008quantum,knez2011evidence,du2015robust}. Up to now, the QSHE has been experimentally realized in monolayer two-dimensional 1T'-WTe$_2$\cite{ok2019custodial,lee2021gate} material systems and quantum well systems, such as HgTe/CdTe\cite{konig2007quantum,nowack2013imaging,deacon2017josephson} and InAs/GaSb\cite{liu2008quantum,knez2011evidence,du2015robust}. In general, the QSHE is accompanied by time-reversal (TR)  symmetry with helical edge states, which is characterized by the topological invariant $Z_2$\cite{kane2005z,fukui2007topological} or spin-Chern number\cite{sheng2005nondissipative,sheng2006quantum,prodan2009robustness}. Ideally, the spin current in QSHE can be transported without any energy dissipation. Although the TR symmetry is considered to be critical to QSHE, it is found that the QSHE can still survive even if the TR symmetry is broken in the presence of an exchange field. The spin-Chern numbers are robust against the exchange field in a wide range, which indicates that the TR symmetry broken QSHE is topologically nontrivial.

Recently, the interesting reentrant phenomena of topological systems and non-Hermitian systems with quite different physical origins have been studied in the literature\cite{goerbig2003microscopic,liu2012observation,beugeling2012reentrant,hsu2013reentrant,yakunin2020unconventional,chen2021evolution,jiang2021mobility,han2022dimerization}. The reentrant integer quantum Hall effect (QHE) is experimentally observed in GaAs quantum wells and theoretically explained\cite{liu2012observation}. The reentrant behavior of quantum anomalous Hall effect (QAHE) is also predicted by applying the magnetic field in Mn-doped HgTe quantum wells, which is due to the competition between the exchange interaction and the direct Zeeman coupling induced by the magnetic field\cite{hsu2013reentrant}. The random disorder induced Berry curvature splitting mechanism can generate the reentrant QAHE in an intrinsic magnetic topological insulator\cite{chen2021evolution}. More recently, the reentrant localization phenomena as the random disorder strength is increased are found and discussed in a non-Hermitian quasiperiodic lattice\cite{wu2021non}. Thus, it is highly desirable to know whether the reentrant phenomena can also appear in QSHE.

In this paper, we demonstrate that the reentrant behavior of QSHE at the Fermi energy can occur when the exchange field is introduced in a two-dimensional topological insulator model system. Through analyzing the band structure and Chern number calculated by the Berry curvature in the momentum space, we find that the Chern number of the lowest energy band for spin-up/spin-down electrons is $\pm1$ while it is the opposite for the corresponding highest energy band. Due to the bulk-edge boundary correspondence, the spin-up and spin-down edge states simultaneously exist in three different bandgap regions. With the presence of an exchange field, the TR symmetry preserved QSHE at Fermi level first becomes TR symmetry broken QSHE. Meanwhile, the QAHE with proper electron energy can be achieved due to the spin splitting mechanism. When the exchange field reaches a critical value, the TR symmetry broken QSHE reappears at the Fermi level. The whole evolution process is clearly presented by studying the energy dependent spin-Chern number. The robustness of TR symmetry broken QSHE and QAHE edge states are further confirmed by calculating the disorder averaged transmission. Finally, we propose a theoretical realization of the tight-binding model in a topolectrical circuit, which can present the reentrant QSHE phenomena with the same frequency by varying the ground inductors.

\emph{Model Hamiltonian}-To start, we first introduce the tight-binding model Hamiltonian based on the $p_{x}$ and $p_{y}$ orbitals with spin degrees of freedom defined on a two-dimensional (2D) honeycomb lattice\cite{zhou2021quantum}, which can describe the underlying physics in several typical two-dimensional materials, for instance, arsenene\cite{zhang2018recent,gui2019two}, antimonene\cite{zhou2015quantum,zhou2016quantum,ji2016two,xu2020optically}, etc. The Hamiltonian can be expressed as
\begin{equation}
\begin{aligned}
H
&= \lambda_{SO}\sum_{i}c_{i}^{\dagger} \sigma_{z} \otimes s_{z} c_{i}+\sum_{i}U_{i} c_{i}^{\dagger} \sigma_{0} \otimes s_{0} c_{i}+\\
&(\sum_{i}\sum_{j=1,2,3}c_{i}^{\dagger} T_{\delta_{j}}c_{i+\delta_{j}}+H.c.)+g\sum_{i}c_{i}^{\dagger} \sigma_{0} \otimes s_{z} c_{i}.\label{HamH}
\end{aligned}
\end{equation}
Here, $c_{i} ^{\dagger}$ and $c_{i}$ represent the creation and annihilation operators on site $i$. $\sigma$ and $s$ indicate the Pauli matrices acting on orbital and spin spaces. The first term describes the intrinsic SOC with coupling strength $\lambda_{SO}$. The second term is the staggered potential with $U_{i} = U(-U)$ for the A(B) sublattice in the system. The third term describes the nearest hopping from site $i$ to $i+\delta_{j}$,
\begin{equation}
T_{\delta_{j}} =
\begin{bmatrix}
t_{1} & z^{(3-j)}t_{2}\\
z^{j}t_{2} & t_{1}
\end{bmatrix}
 \otimes s_{0} \label{HamT}
\end{equation}
where $z = e^{2i\pi/3}$ and $t_{1/2}$ is the hopping coefficient. When $\lambda_{SO} < U$, the system is in quantum valley Hall effect (QVHE) phase. When $\lambda_{SO} > U$, the system is in QSHE phase\cite{zhou2021quantum}. The last term represents a uniform exchange field introduced in the system with strength $g$\cite{yazyev2010emergence}. In the following, we focus on the $\lambda_{SO} > U$ case. The  Fermi level is assumed set as $E=0$ and parameters of Hamiltonian are set as $t_{1} = 1$, $t_{2} = 1$, $\lambda_{SO} = 0.44$, $U = 0$ without loss of generality.

\begin{figure}
\centering
\includegraphics[scale=0.39]{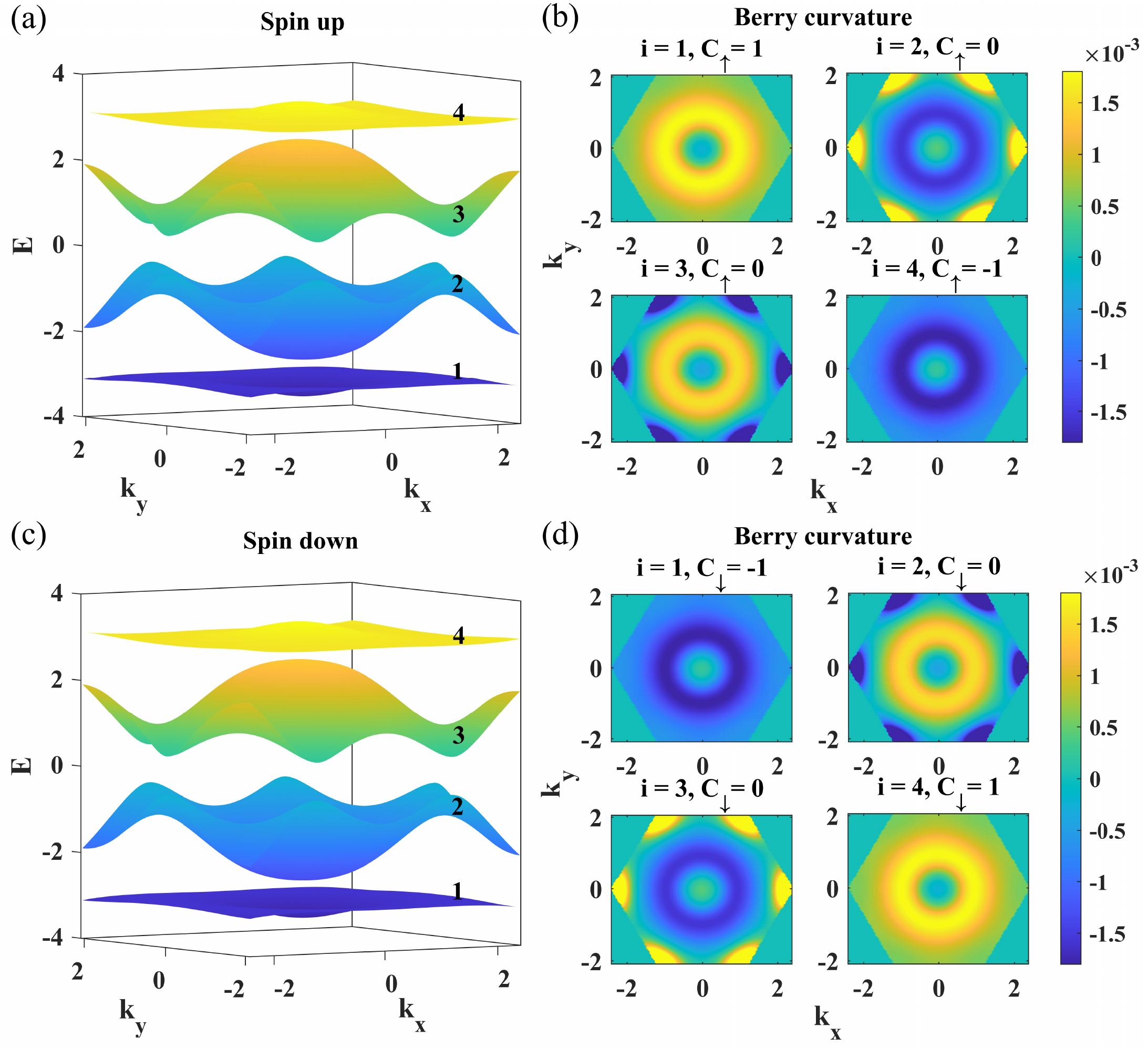}
\caption{(a,c) The band structure in momentum space for spin-up and spin-down components, respectively. Each band is denoted as a number from the bottom to top marked on the right. (b,d) The corresponding Berry curvature distribution in momentum space and Chern number of each band for spin-up and spin-down components, respectively.}
\label{Fig1}
\end{figure}

Before analyzing the exchange field effect on the QSHE, we briefly discuss the bulk electronic property by taking the Fourier transform of the tight-binding Hamiltonian into the momentum space,
\begin{equation}
\begin{aligned}
H(\vec{k})
&=\begin{bmatrix}
H_{11}(\vec{k})& H_{12}(\vec{k})\\ H_{21}(\vec{k}) & H_{22}(\vec{k})
\end{bmatrix},
\end{aligned}
\end{equation}
where
\begin{equation}
\begin{split}
H_{11}(\vec{k}) &= \lambda_{SO}\sigma_{z} \otimes s_{z}+U\sigma_{0} \otimes s_{0}+g \sigma_{0}\otimes s_{z},\\
H_{22}(\vec{k}) & =\lambda_{SO}\sigma_{z} \otimes s_{z}-U\sigma_{0} \otimes s_{0}+g \sigma_{0}\otimes s_{z},\\
H_{12}(\vec{k}) &= e^{-i\vec{k}\cdot\vec{e}_{1}}T_{\vec{e}_{1}}+e^{-i\vec{k}\cdot\vec{e}_{2}}T_{\vec{e}_{2}}+e^{-i\vec{k}\cdot\vec{e}_{3}}T_{\vec{e}_{3}},\\
H_{21}(\vec{k}) &= H_{12}^{\ast}(\vec{k}).\notag
\end{split}
\end{equation}
Here the basis vectors are chosen as $\vec{e}_{1} = (0,a), \vec{e}_{2} = (-\frac {\sqrt{3}}{2}a,-\frac {a}{2}), \vec{e}_{3} = (\frac {\sqrt{3}}{2}a,-\frac {a}{2})$.
Figures. \ref{Fig1}(a) and \ref{Fig1}(c) present the band structure of spin-up and spin-down components in momentum space when there is no exchange field. We can clearly see that there are four bands in total, which are well separated in energy. Note that each band is doubly degenerated due to the presence of TR symmetry. In order to know the topological property of each spin-up and spin-down band, we calculate the Berry curvature distribution $\Omega(k_x,k_y)$ in momentum space by using an efficient method in Ref. \onlinecite{fukui2005chern}. As shown in Figs. \ref{Fig1}(b) and \ref{Fig1}(d), the Berry curvature of each band has a ring-shaped distribution near the $\Gamma$ point. For the topological trivial band, the Berry curvatures are also peaked at the corners of the Brillouin zone with an opposite sign to that in the ring region, which results in a net zero Chern number. By furthermore integrating the Berry curvature, the Chern numbers for each band with the different spin components are obtained, i.e., $C_{\uparrow} = (1,0,0,-1)$, $C_{\downarrow} = (-1,0,0,1)$, from which we can find that the lowest and highest energy bands have a nontrivial topological property and the Chern numbers for different spin components are just opposite. Since the spin-Chern number is defined as $C_s=(C_{\uparrow}-C_{\downarrow})/2=(1,0,0,-1)$, the system can realize the QSHE as long as the electron energy lies in three bandgaps.

\begin{figure*}
\centering
\includegraphics[scale=0.6]{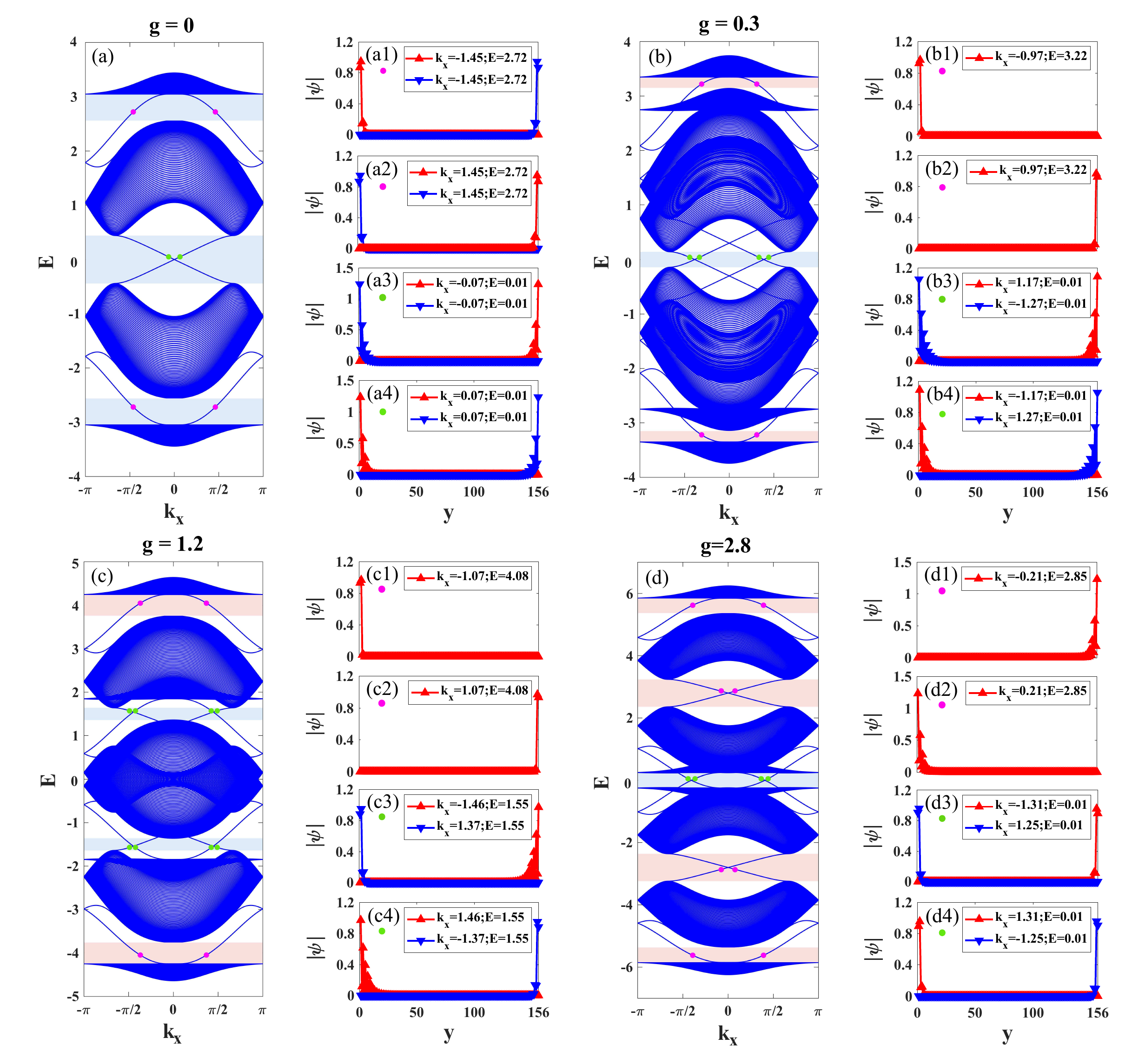}
\caption{(a-d) Band structure of a finite nanoribbon system along the $x$-direction with several exchange field strengths, $g=0, 0.3, 1.2, 2.8$, respectively. The width of the system is $W=156$ and the periodic boundary condition in the $x$-direction. (a1-a4) The absolute value of wave functions with energy $E = 2.72$ and $E = 0.01$ along the transverse $y$-direction for two QSHE denoted as pink and green solid dots in (a). (b1-b4) The absolute value of wave functions with energy $E = 3.22$ for QAHE and $E = 0.01$ for time-reversal (TR) symmetry broken QSHE. (c1-c4) The absolute value of wave functions with energy $E = 4.08$ for QAHE and $E = 1.55$ for TR symmetry broken QSHE. (d1-d4) The absolute value of wave functions with energy $E = 2.85$ for QAHE and $E = 0.01$ for TR symmetry broken QSHE. The red (blue) upper (lower) triangles represent spin-up (spin-down) respectively.}
\label{Fig2}
\end{figure*}
\emph{Finite nanoribbon system}-Without loss of generality, we consider a honeycomb lattice system along the zigzag direction ($x$-direction) with finite width ($W=156$) in the next. A schematic diagram of the structure is shown in the Supplemental Material\cite{suppsee}. First, due to the bulk-edge correspondence, topologically protected edge states with spin degeneracy appear in three bandgap regions (denoted as the light blue region) when $g=0$ as shown in Fig. \ref{Fig2}(a). Note that the chirality of edge states with different spin components in three bandgap regions is determined by the Chern number of its corresponding lowest energy band. When the electron energy is $E = 2.72$ located in the top bandgap region, the absolute value of the wave function of the edge states versus the transverse $y$-direction (marked by the pink dots in Fig. \ref{Fig2}(a) in the top bandgap) are presented in Figs. \ref{Fig2}(a1) and \ref{Fig2}(a2). We observe that the spin-up (spin-down) electrons are moving anticlockwise (clockwise) denoted with red (blue) upper (lower) triangles by combining with group velocity analysis, which is consistent with $C_{\uparrow/\downarrow}=\pm1$ obtained by the Berry curvature calculations. As the electron energy $E$ goes to $0.01$ in the middle bandgap region, the spin-up (spin-down) edge states with positive velocity are mainly distributed in the lower (upper) edge, while the corresponding edge states with negative velocity are mainly distributed in the upper (lower) edge, which are shown in Figs. \ref{Fig2}(a3) and \ref{Fig2}(a4). Detailed information on the wave function when the electron energy is $E = -2.72$ located in the bottom bandgap region is given in the Supplemental Material\cite{suppsee}. Through an edge state analysis, we can know that the QSHE indeed appears in these three bandgap regions as discussed above.

Next, we consider the cases of $g \neq 0$, where a uniform exchange potential with magnitude $g$ $(-g)$ is added to the spin-up (spin-down) component in the Hamiltonian. Since spin degeneracy is lifted with the presence of an exchange field, the spin-up bands move upward and the spin-down bands move downward at the same time. As an example, the band structure with strength $g=0.3$ is shown in Fig. \ref{Fig2}(b). Figures. \ref{Fig2}(b1)-\ref{Fig2}(b4) present the distribution of the absolute value of the wave functions when the electron energy is located in the top and middle bandgap regions with $E = 3.22$ and $E = 0.01$. In the top bandgap region, only the spin-up chiral edge states are preserved and give rise to the spin-polarized QAHE (light pink region in Fig. \ref{Fig2}(b)). Correspondingly, the spin-down edge states are left in the bottom bandgap region, which gives out another spin-polarized QAHE region (light pink region in Fig. \ref{Fig2}(b)), while for the middle bandgap region, the spin-up and spin-down edge states with opposite chirality still persist, which produces a TR symmetry broken QSHE (light blue region in Fig. \ref{Fig2}(b)). It is worth mentioning that the TR symmetry broken QSHE and the QAHE coexist in the system. The transition from the TR symmetry broken QSHE to the QAHE can actually be achieved by tuning the electron energy via a gate voltage.

When the exchange field strength $g$ continues to increase up to $1.2$, four bandgap regions appear in total and the bandgap in the middle region around the Fermi level is closed as shown in Fig. \ref{Fig2}(c). The spin-up and spin-down bulk bands mix together to form a metallic state around the Fermi level. By investigating the wave function distributions when $E = \pm 1.55$ in the two new bandgap regions (see Figs. \ref{Fig2}(c3) and \ref{Fig2}(c4)), we can know that the TR symmetry broken QSHE states are generated. At the same time, the QAHE states in the top and bottom bandgap regions are still preserved (see Figs. \ref{Fig2}(c1) and \ref{Fig2}(c2)). Note that the edge states in the bottom bandgap correspond to spin-down electrons. Thus, four bandgap regions correspond with two TR symmetry broken QSHE and two QAHE.

In the following, we discuss the situation with a larger exchange field to realize the reentrant behavior of QSHE at the Fermi energy. As shown in Fig. \ref{Fig2}(d), the number of bandgap regions is further increased to five. Interestingly, the middle bandgap region is reopened compared with that in Fig. \ref{Fig2}(c). According to the wave function analysis in Figs. \ref{Fig2}(d3) and \ref{Fig2}(d4), we can know that the TR symmetry broken QSHE at the Fermi energy $E_{F} = 0.01$ reenters. This reentrant behavior of QSHE is purely induced by the exchange field. Since the spin-up (spin-down) bands are shifted upwards (downwards), the other four bandgap regions are spin-polarized QAHE. As an illustration, the absolute value of the wave functions corresponding to the spin-up edge states when electron energy $E = 2.85$ are shown in Figs. \ref{Fig2}(d1) and \ref{Fig2}(d2). Due to the presence of the spin $S_z$ conservation, the edge states in Fig. 2(b)-2(d) are not gapped out. Other detailed information on the wave function distributions is provided in the Supplemental Material\cite{suppsee}. When the exchange field strength $g$ is greater than 3, the spin-up and spin-down bands are completely separated and the system becomes a normal insulator.

\begin{figure}
\centering
\includegraphics[scale=0.37]{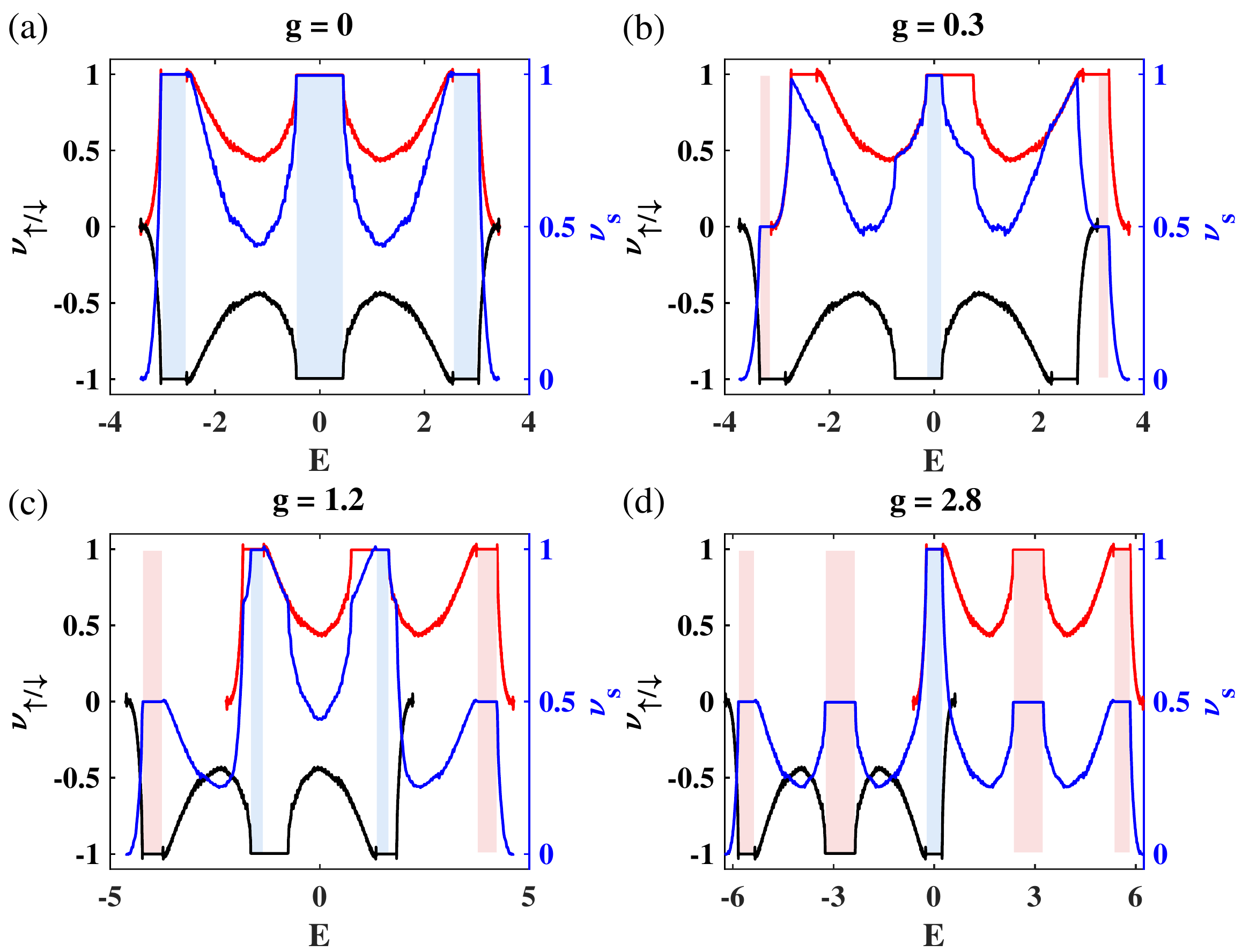}
\caption{The energy dependent Chern number for spin-up/down component with red/black curves in the left axis and energy dependent spin-Chern number $\nu_{s}$ with blue curve in the right axis vs the electron energy under different exchange field strengths. (a) $g = 0$, (b) $g = 0.3$, (c) $g = 1.2$ and (d) $g = 2.8$.}
\label{Fig3}
\end{figure}

\emph{Energy dependent spin-Chern number calculations}-In this section, we verify the topological property of the finite system by introducing the energy dependent  spin-Chern number $\nu_s(E)=\frac{1}{2}(\nu_{\uparrow}(E,r_0)-\nu_{\downarrow}(E,r_0))$ where $\nu_{\uparrow/\downarrow}$ is the energy dependent Chern number for the spin-up/spin-down component calculated within the center of the system $r_0$, which can be computed by the antisymmetric product of the projection operators\cite{cheng2021antichiral,mitchell2018amorphous,kitaev2006anyons,jiang2019experimental,suppsee}. In the numerical simulation, the system size is fixed as $30 \times 156$ sites. The calculated results of $\nu_{\uparrow/\downarrow,s}$ versus the electron energy with different exchange fields are given in Fig. \ref{Fig3}. Without an exchange field, $\nu_{\uparrow}$ and $\nu_{\downarrow}$ are $\pm1$ in the corresponding three bandgap regions in Fig. \ref{Fig2}(a), resulting in the formation of the QSHE with $\nu_{s}=1$. As the exchange field strength increases, $\nu_{\uparrow}$ and $\nu_{\downarrow}$ shift in opposite directions along the electron energy axis. In Fig. \ref{Fig3}(b), one TR symmetry broken QSHE and two spin-polarized QAHE are formed with different electron energies. When the exchange field strength $g=1.2$, four topological regions appear with two QSHE and two QAHE regions as shown in Fig. \ref{Fig3}(c). More importantly, with the exchange field strength further increasing, the opposite Chern number plateaus at the Fermi level meet again, which indicates the reentrant phenomena of the QSHE. Note the energy dependent spin-Chern number calculations are consistent with the analysis of the band structure and wave function distribution of edge states in Fig. \ref{Fig2}. To further verify the robustness of the generated QSHE and QAHE, the transport properties of the corresponding edge states in the presence of random disorders are presented in the Supplemental Material\cite{suppsee}(see, also, Refs. \cite{datta1997electronic,xing2011topological,zhang2014universal,khomyakov2005conductance,wang2009relation,sorensen2009efficient,zhang2012first}therein).

\begin{figure}
\centering
\includegraphics[scale=0.59]{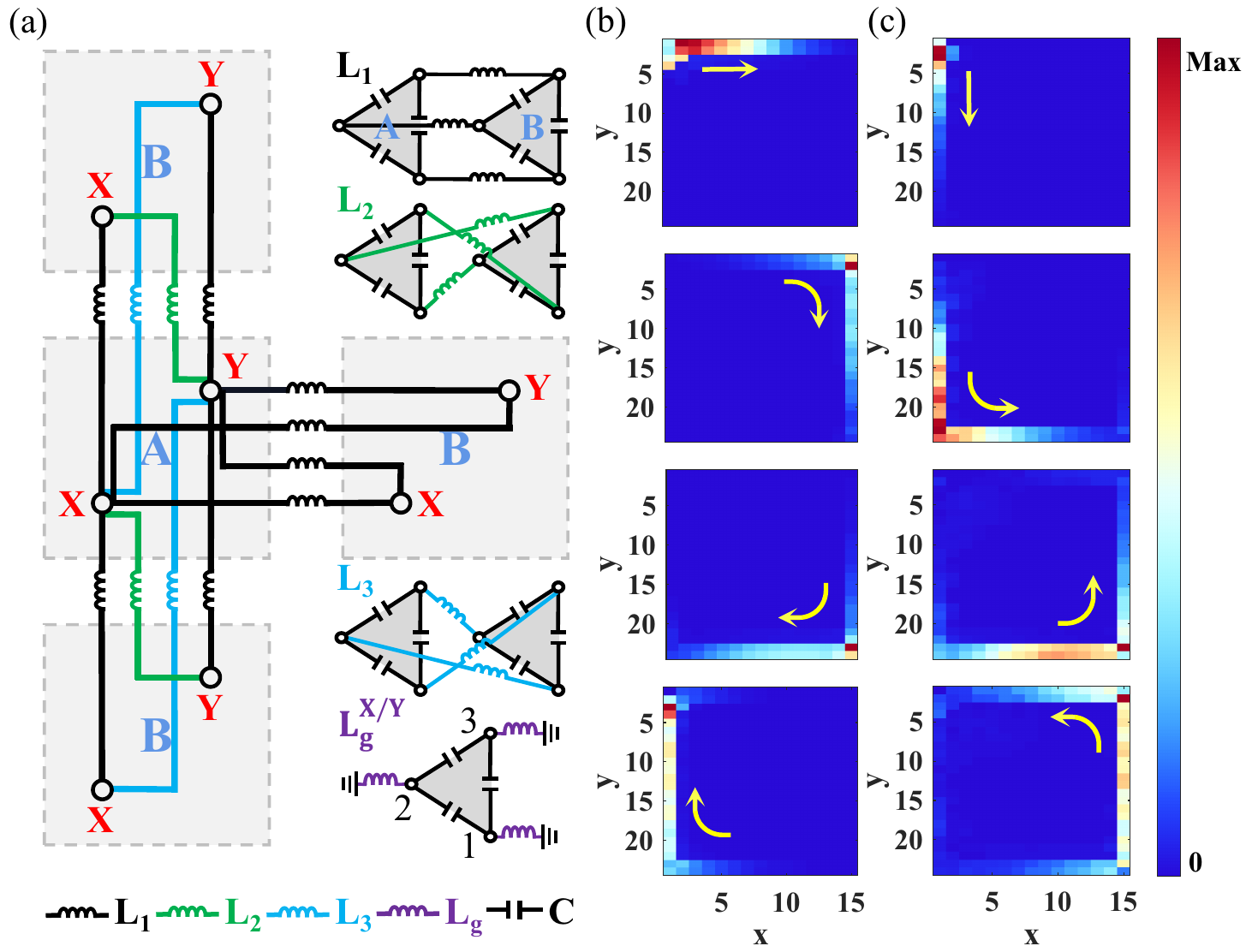}
\caption{(a) Detailed circuit configuration for the A site and its connections to three nearest neighbor B sites in a topolectrical circuit. Each A/B site consists of two nodes (X/Y) to simulate $p_{x/y}$ orbital. In each node, there are three sub-nodes with grounded inductors ($L^{X,Y}_{g,i},i=1,2,3$) to simulate spin degrees of freedom, which are connected by capacitors. Three different types of connections from the A site to B sites are denoted by black, green and blue lines. (b,c) Time evolution of the Gaussian wave packets excited with a pseudospin-up/pseudospin-down source, respectively. Four subplots (top to bottom) of each column indicate the trajectory of the wave packet in turn with time points $t_{0}, 2.5t_{0}, 4t_{0}, 5.5t_{0}$, respectively. Here, $L_{1}=L_{2}=L_{3}=1, C=1,L^X_{g,1}=L^Y_{g,1}=0.4,L^X_{g,2}=0.11,L^X_{g,3}=0.4,L^Y_{g,2}=0.12,L^Y_{g,3}=0.3$. }\label{Fig4}
\end{figure}

\emph{Topolectrical circuit implementation}-Topolectrical circuits (TECs) are powerful platforms for realizing large numbers of exotic topological physics\cite{imhof2018topolectrical,lee2018topolectrical,zhao2018topological,helbig2019band,zhu2018simulating,albert2015topological,zhang2019topological,zhang2021experimental,ningyuan2015time,albert2015topological,lee2018topolectrical,haenel2019chern,ezawa2019electric,hofmann2019chiral,dong2021topolectric}. Here, we design a TEC to realize the reentrant phenomena of QSHE. An equivalent TEC with capacitors and inductors to simulate the tight-binding Hamiltonian in Eq. (\ref{HamH}) is presented in Fig. \ref{Fig4}(a). To realize the reentrant phenomena of QSHE, we design a finite TEC system composed of $15 \times 6$ unit cells and numerically calculate the time evolution of the Gaussian wave packet. Here the wave packet with the fixed frequency $\omega_0=1.55$ in the middle bandgap is excited by the pseudospin-up/pseudospin-down source corresponding with the exchange field $g = 2.8$ case. As shown in Figs. \ref{Fig4}(b) and \ref{Fig4}(c), we can see that the wave packets with pseudospin-up/pseudospin-down propagate along the system boundary in the anticlockwise and clockwise directions, which indicates that the reentrant QSHE is formed. Detailed information about TEC and wave packet dynamics with $\omega_0=1.55$ and $g=0$ case can be found in the Supplemental Material\cite{suppsee}.

\emph{Conclusions}-We realize the reentrant phenomena of the TR symmetry broken QSHE at the Fermi level by introducing a uniform exchange field to a four-band topological insulator. Moreover, as the exchange field strength increases, the QSHE and spin-polarized QAHE can simultaneously exist in the system. The topological properties and robustness of the generated QSHE and QAHE are verified by calculating the energy dependent spin-Chern numbers and the disorder averaged transmission coefficients $\langle T \rangle$. Finally, the realization of reentrant phenomena of QSHE is discussed in a topolectrical circuit system. Our work extends the reentrant phenomena into the QSHE regime and discusses their physical realization.

We gratefully acknowledge the support from the National Natural Science Foundation of China (Grant No. 12074230, 12174231, 11974355, 12147215), the Fund for Shanxi ``1331 Project", Shanxi Province 100-Plan Talent Program, Fundamental Research Program of Shanxi Province through 202103021222001. This research was partially conducted using the High Performance Computer of Shanxi University.

\bigskip

\noindent{$^{*)}$chenjun@sxu.edu.cn}\\
\noindent{$^{\dagger)}$zhanglei@sxu.edu.cn}

\bibliography{ref}

\end{document}